\documentclass[aps,prd,twocolumn,showpacs,groupedaddress,superscriptaddress,amsmath,amssymb,floatfix]{revtex4-2} 

\usepackage[utf8]{inputenc}
\usepackage[english]{babel}										
\usepackage{bbm}
\usepackage[T2A]{fontenc}

\newcommand{\G}{{\text{G}}}
\newcommand{\DM}{{\text{DM}}}

\newcommand{\vect}[1]{\boldsymbol{#1}}

\def\nn{\nonumber}

\usepackage{color}			
\usepackage{graphicx}
\usepackage{bm}

\usepackage{physics}
\usepackage{braket}
\usepackage{slashed}

\usepackage[colorlinks=true, filecolor=blue, citecolor=blue,urlcolor=blue]{hyperref}
\usepackage{url}
\begin{document}
	
\abovedisplayskip = 4pt
\belowdisplayskip = 4pt
\abovedisplayshortskip = 4pt
\belowdisplayshortskip= 4pt

\title{The bremsstrahlung-like production of the massive spin-2 dark matter mediator} 

\author{I.~V.~Voronchikhin}
\email[\textbf{e-mail}: ]{i.v.voronchikhin@gmail.com}
\affiliation{ Tomsk Polytechnic University, 634050 Tomsk, Russia}

\author{D.~V.~Kirpichnikov}
\email[\textbf{e-mail}: ]{dmbrick@gmail.com}
\affiliation{Institute for Nuclear Research, 117312 Moscow, Russia}

\begin{abstract}
The link between Standard Model (SM) particles and  dark matter (DM)
can be introduced via spin-2 massive mediator, $ \G$, that couples to photon and charged leptons. Moreover, in a mediator mass
range from sub-MeV to sub-GeV, fixed-target facilities such as NA64e, LDMX, NA64$\mu$,  M$^3$, and E137, can
potentially probe such particle of the hidden sector via the  signatures that are described
by the bremsstrahlung-like process involving tensor mediator. We  compare numerically
the Weizsaker-Williams (WW) approximation and the exact tree-level (ETL) approach for the bremsstrahlung-like 
mediator  production cross section by choosing various parameters of the fixed-target experiments. 
In addition, we derive novel constraints on  spin-2 DM mediator parameter space
from the data of the  E137 fixed-target experiment.  In particular, we demonstrate  that the E137 experiment has been ruled out the 
the couplings of the spin-2  mediator at the level of~$8\times10^{-8}~\mbox{GeV}^{-1}~\lesssim~c^\G_{ee}~\lesssim~10^{-5}~\mbox{GeV}^{-1}$
for the typical  masses in the range   $100~\mbox{MeV}~\lesssim~m_\G~\lesssim 800~\mbox{MeV}$, 
that corresponds to the statistics of $1.87\times 10^{20}$ electrons 
accumulated on target. The latter implies its universal  coupling to
photons and leptons, $c^{\rm G}_{ee} = c^{\rm G}_{\gamma \gamma}$. 
\end{abstract}

\maketitle

\section{Introduction}

During the past few decades, a considerable number of astrophysical observations have led to the concept of 
the dark sector, where the DM can  manifest itself via gravitational 
effects~\cite{Bergstrom:2012fi,Bertone:2016nfn}. 
In particular, the main indirect evidence of the DM is the galaxy rotation velocities, the cosmic microwave 
background anisotropy, and the gravitational lensing~\cite{Cirelli:2024ssz,Bertone:2004pz,Gelmini:2015zpa}.
Also, the DM can be exploited as a solution of well-known problems of modern physics~(e.g., the anomalous 
magnetic moment puzzle~\cite{Aoyama:2020ynm}, large-scale structures~\cite{Davis:1985rj}, etc). 
Moreover, about $\simeq 85\%$ the total amount of matter in our Universe is approximately invisible, 
although the fundamental particle nature of DM still remains the great 
mystery~\cite{Planck:2015fie,Planck:2018vyg}.
Thus, the assumption of the dark matter existence motivates the development of SM extensions.

The interaction between SM particles and sub-GeV DM
in the thermal bath of the early Universe leads to overproduction of DM particles 
that require a depletion mechanism to yield the observed density of
DM~\cite{Lee:1977ua,Kolb:1985nn,Krnjaic:2015mbs}.
Specifically,  an additional connection between DM and SM particles via mediator~(MED) as a new portal can  lead to observed relic abundance of DM.
For example, typical scenarios of DM and the corresponding thermal targets involving dark boson 
mediators encompass particles with spin-0~\cite{McDonald:1993ex,Burgess:2000yq,Wells:2008xg,Schabinger:2005ei,Bickendorf:2022buy,Boos:2022gtt,Sieber:2023nkq}, spin-1~\cite{Catena:2023use,Holdom:1985ag,Izaguirre:2015yja, Essig:2010xa, Kahn:2014sra,Batell:2014mga,Izaguirre:2013uxa,Kachanovich:2021eqa,Lyubovitskij:2022hna,Gorbunov:2022dgw,Claude:2022rho,Wang:2023wrx}, and spin-2~\cite{Lee:2013bua,Kang:2020huh,Bernal:2018qlk,Folgado:2019gie,Kang:2020yul,Dutra:2019xet,Clery:2022wib,Gill:2023kyz,Wang:2019jtk,deGiorgi:2021xvm,deGiorgi:2022yha,Jodlowski:2023yne}.

Current and planned fixed-target experiments can provide the constraints on Dirac and scalar  DM~\cite{Voronchikhin:2022rwc},  implying the effective interaction between the 
tensor mediator and (dark)  matter. 
In the present paper, we fill the gap in the literature dedicated to the bremsstrahlung-like 
production of spin-2 hidden particle in the lepton fixed-target experiments
In particular, we study  the implication of the WW approximation for calculating the cross section 
of tensor-mediator production in various accelerator-based experiments.

To be more specific, we continue the study of 
authors of 
Refs.~\cite{Gninenko:2017yus,Liu:2016mqv,Liu:2017htz,Kirpichnikov:2021jev,Sieber:2023nkq,Voronchikhin:2024vfu} and  
calculate the production cross sections of spin-2 DM mediator by using the exact-tree-level (ETL) and WW methods 
for NA64e, NA64$\mu$, LDMX, and $\mbox{M}^3$ experiments. 
The implication of the WW 
approximation~\cite{Fermi:1925fq,vonWeizsacker:1934nji,Williams:1935dka,Budnev:1975poe} for spin-0
and  spin-1 MED has been investigated for: (i) a electron mode of the E137 
experiment~\cite{Liu:2016mqv,Liu:2017htz}, (ii) a muon mode in the case of 
NA64$\mu$~\cite{Kirpichnikov:2021jev,Sieber:2023nkq}, and (iii) a proton 
bremsstrahlung~\cite{Blumlein:2013cua,Foroughi-Abari:2021zbm,Foroughi-Abari:2024xlj,Gorbunov:2023jnx,Harland-Lang:2019zur,Gorbunov:2024vrc,Gorbunov:2024iyu}.

In addition, we calculate the experimental sensitivity of E137 and LDMX involving visible decays of spin-2 
DM mediator in the mass range of interest, $100~\mbox{MeV}~\lesssim~m_\G~\lesssim~1~\mbox{GeV}$. That benchmark 
simplified  scenario implies the universal  coupling of spin-2 particle with  the charged leptons and photon.

Our study might be complementary to Refs.~\cite{Jodlowski:2023yne,dEnterria:2023npy}, that explicitly 
discussed  the  accelerator-based bounds on photon-specific spin-2 DM mediator. 
We note that authors of Refs.~\cite{Folgado:2019gie,Kang:2020huh,Lee:2024wes} consider explicitly the  
ultra violet  completion for the  scenarios of spin-2 DM mediator in the context of the warped extra 
dimension and discuss the thermal production mechanisms for dark matter with various channels and the dark 
matter self-scattering. The latter analysis is beyond the scope of the present paper, in our study
we just adopt the DM relic abundance curves from~\cite{Kang:2020huh} for vector, scalar and Dirac DM,
that are complementary to the  sensitivity of~E137, implying simplified massive spin-2 DM mediator scenario.

In Sec.~\ref{sec:BenchModels} 
we discuss the benchmark models of the \text{spin-2} mediator
and main parameters of the considered lepton fixed-target experiments.
In Sec.~\ref{sec:BremsstrahlungLike} we briefly describe the procedure of cross section calculation in 
the case  of bremsstrahlung-like production of mediator at lepton fixed-target experiments and 
derive the corresponding  matrix elements. In Sec.~\ref{sec:crossSectionsBremssLike} 
we discuss the differential and total cross sections for the tensor mediator and lepton fixed-target 
experiments.   
In Sec.~\ref{sec:VisModConstr} we discuss the number of signal events and derive the experimental reach of E137 and LDMX in the visible mode.
We conclude in Sec.~\ref{sec:Conclusion}. In Appendices~\ref{FPaction} and~\ref{sec:MatEl2to3Graviton}
we collect some helpful formulas. 

\section{Benchmark scenarios and experiments \label{sec:BenchModels}}

\subsection{Model of mediator}

Effective Lagrangian density of the massive spin-2 mediator and matter reads as~\cite{Lee:2013bua,Folgado:2019sgz,Kang:2020huh}:
\begin{equation}
\mathcal{L}_{\rm eff}^{\rm G} 
\supset    
- 
\sum_{i}
c^{\rm G}_{i} 
G^{\mu \nu} {T^{i}}_{\mu \nu},
\end{equation}
where~${T}^i_{\mu \nu}$ is the energy-momentum tensor (EMT) 
and~$c^{\rm G}_{i}~=~(c^{\rm G}_{\rm DM}, c^{\rm G}_{\rm SM})$ is a coupling constant of dimension 
$\mbox{GeV}^{-1}$ between the massive spin-2 mediator and corresponding 
types of (dark) matter.

In the cases of scalar,~$S$, vector,~$V^{\mu}$, and Dirac 
fermion,~$\psi$, types of massive DM particles, the symmetrized EMTs  read, 
respectively~\cite{Kang:2020huh}:
\begin{multline}
    T_{\mu \nu }^{\rm S} 
= 
        \partial_\mu S \partial_\nu S
    -   (1/2) \eta_{\mu \nu}
        \left[
            (\partial_\rho S )^2
        -   m_{\rm S}^2 S^2
        \right],  
\end{multline}
\begin{multline}
    T_{\mu \nu }^{\rm V} 
= 
    (1/4) \eta_{\mu \nu} V_{\lambda \rho} V^{\lambda \rho} 
    - V_{\mu \lambda} {V^{\lambda}}_{ \nu}
    + m_{\rm V}^{2} V_{\mu} V_{\nu} 
- \\ -
    (1/2) m_{\rm V}^{2} \eta_{\mu \nu} V_{\alpha} V^{\alpha},  
\end{multline}
\begin{multline}
    T_{\mu \nu }^\psi 
= 
    - (i/4)
    \bar{\psi} \left[ 
      \gamma_\mu \overleftrightarrow{\partial}_\nu 
    +  \gamma_\nu \overleftrightarrow{\partial}_\mu  
    \right] \psi
+ \\ + 
    \eta_{\mu \nu}m_{\psi} \bar{\psi} \psi
    +  (i/2) \eta_{\mu \nu}    
    \bar{\psi} \gamma^\rho \overleftrightarrow{\partial}_\rho \psi,
\end{multline}
where we denote 
$
	\overline{\psi} \overleftrightarrow{\partial}_{\mu} \psi
= 
		\left( \partial_{\mu} \overline{\psi} \right) \psi 
	- 	\overline{\psi} \left( \partial_{\mu} \psi \right)
$ and $m_{\rm DM}~=~(m_{\rm S},~m_{\rm V},~m_{\psi})$ is a  mass of the corresponding DM type.
Next, the EMT of SM particles reads as~\cite{Folgado:2019sgz}:
\begin{multline}
    T_{\mu \nu }^{\rm SM} 
=
    \left[
    \frac{1}{4} \eta_{\mu \nu} F_{\lambda \rho} F^{\lambda \rho} 
    -   F_{\mu \lambda} {F^{\lambda}}_{ \nu}
    \right]
    - \\ -
    	\frac{i}{4}
    	\overline{l}
    	\left[  \gamma_\mu \overleftrightarrow{D}_\nu +  \gamma_\nu \overleftrightarrow{D}_\mu  \right]l
    + \frac{i}{2} 
    \eta_{\mu \nu} \overline{l} \gamma_{\rho} \overleftrightarrow{D}^{\rho} l,
\end{multline}
where~$l$~ is the SM lepton, $F_{\mu\nu}~=~\partial_\mu A_\nu~-~\partial_\nu A_\mu$~ is the strength tensor for the SM photon field~$A_\mu$, $D_\mu~=~\partial_\mu~-~ieA_\mu$ is the covariant derivative for the~$U(1)$ gauge field, we use notation~$
	\overline{l} \overleftrightarrow{D}_{\mu} l 
= 
		\left( D_\mu^* \overline{l} \right) l 
	- 	\overline{l} \left( D_\mu l \right).
$

\begin{figure}[!tb]
\centering
\includegraphics[width=0.49\textwidth]{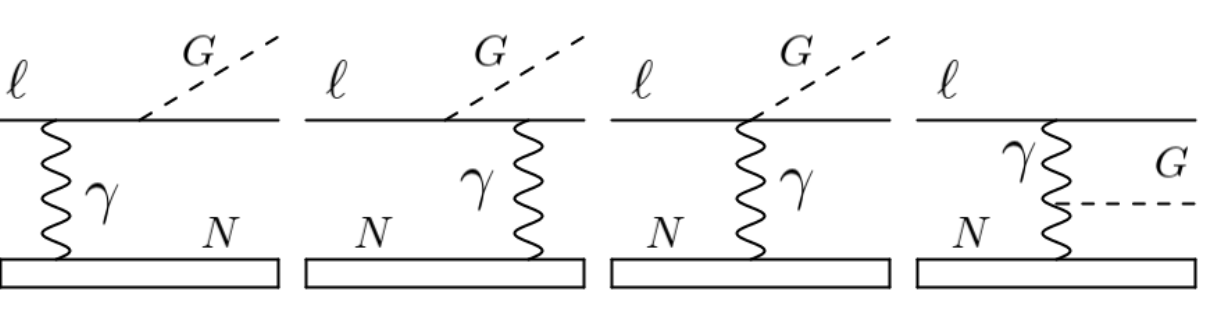}
\caption{Feynman diagrams describing bremsstrahlung-like   signature for the tensor mediator.
Typically, one can consider two benchmark scenarios for spin-2 DM mediator~\cite{Kang:2020huh}.  
It can decay either to the pair of SM particles, $\mbox{G}\to \mbox{SM}+\mbox{SM}$, or to the invisible DM sector,
$\mbox{G}\to \mbox{DM}+\mbox{DM}$, 
implying its branching fraction is $\mbox{Br}(\mbox{G}\to \mbox{DM}+\mbox{DM})\simeq 1$. 
\label{eNToeNGDiagram} }
\end{figure}

The spin-2 vertices involving fermion, vector and scalar field in the momentum space read as, respectively:
\begin{align}
    T_{\mu \nu }^{({ \rm G} f f)}(p_1, p_2)      \label{eq:FeynRuleGravitonDirac}
& =
    (-ic_{ff}^{\rm G})/4
    \Big[
    \gamma_{\mu}({p_1} + {p_2} )_\nu
    +   \gamma_{\nu}({p_1} + {p_2})_\mu
  \nn \\ &  -
    2 \eta_{\mu \nu} ( \slashed{p}_1 + \slashed{p}_2 - 2 m_{f} )
    \Big],
\\
    T_{\mu \nu \alpha \beta }^{(\rm G V V)}(k_1, k_2) \label{eq:FeynRuleGravitonVector}       
& =   
 (-ic_{\rm V V}^{\rm G}) \Bigl[   m_{\rm V}^{2} 
    C_{\mu \nu \alpha \beta}
+ \nn \\ & +
    \left(
    C_{\mu \nu \alpha \beta \sigma \lambda} 
    +   C_{\nu \mu \alpha \beta \sigma \lambda}
    \right) {k_1}^{\sigma}{k_2}^{\lambda} \Bigr],
\\
    T_{\mu \nu }^{(\rm G S S)}(q_1, q_2)      \label{eq:FeynRuleGravitonScalar} 
& =
      (-ic_{SS}^{\rm G}) \left[ \eta_{\mu \nu} m_{\rm S}^2 
    -   {q_1}^{\alpha} {q_2}^{\beta} C_{\mu \nu \alpha \beta} \right],
\end{align}
where $p_1,~p_2$ are  4-momenta of fermions, $k_1,~k_2$~are \text{4-momenta} of vector particles $V_{\alpha}(k_1)$ and $V_{\beta}(k_2)$, $q_1,~q_2$~are 4-momenta of scalar particles, in Eqs.~(\ref{eq:FeynRuleGravitonVector})  and~(\ref{eq:FeynRuleGravitonScalar})
we use the following notations:
\begin{equation*}
    C_{\mu \nu \alpha \beta}
=
    \left(   \eta_{\mu \alpha} \eta_{\nu \beta} 
    +   \eta_{\nu \alpha} \eta_{\mu \beta} 
    -   \eta_{\mu \nu} \eta_{\alpha \beta}
    \right),
\end{equation*}
\begin{multline*}
    C_{\mu \nu \alpha \beta \sigma \lambda}
=
    \frac{1}{2} \eta_{\mu \nu} 
    (   \eta_{\beta \sigma} \eta_{\alpha \lambda} 
    -   \eta_{\sigma \lambda} \eta_{\alpha \beta} )
    +   \eta_{\alpha \beta} \eta_{\mu \sigma} \eta_{\nu \lambda}
- \\ -   
        \eta_{\mu \beta} \eta_{\nu \sigma} \eta_{\alpha \lambda}
    +   \eta_{\mu \alpha} 
    (   \eta_{\beta \nu} \eta_{\sigma \lambda} 
    -   \eta_{\sigma \beta} \eta_{\lambda \nu} ).
\end{multline*}
In addition, the 4-legs vertex reads:
\begin{equation}
    T_{\mu \nu \alpha}^{({ \rm G} l l \gamma)}  \label{eq:FeynRuleGravitonPhotonLepton} 
= 
    (-e/2) (-ic_{ll}^{\rm G})
    \left[
    \eta_{\alpha \nu} \gamma_{\mu}
    +   \eta_{\mu \alpha} \gamma_{\nu}
    -   2 \eta_{\mu \nu} \gamma_{\alpha}
    \right].
\end{equation}
The Feynman diagrams are shown in Fig.~\ref{eNToeNGDiagram} for the bremsstrahlung-like production of spin-2 mediator.

\begin{table}[tb]
    \begin{tabular}{lccccc}
	\hline
 \hline
	& NA64e  & LDMX & NA64$\mu$  & M$^3$ & E137 \\ \hline
        target material & Pb & Al & Pb &  W & Al \\ \hline
	$Z$,~\mbox{atomic number} & $82$ & $13$ & $82$ &  $74$ & $13$  \\ \hline
	$A,~\mbox{g}\cdot\mbox{mole}^{-1}$  & $207$ & $27$ & $207$ &  $184$ & $27$ \\ \hline
	$x_{\rm cut}=E^{\rm cut}_{\rm G}/E_l$ & $0.5$ & $0.7$ & $0.5$ &  $0.4$  & $0.1$ \\ \hline
	$l^\pm$,  primary beam & $e$ & $e$ & $\mu$ &  $\mu$ & $e$ \\ \hline
        $E_l$,~GeV,~\mbox{beam energy} & $100$ & $16$ & $160$ &  $15$  & $20$ \\  \hline
        vis. mode, $\mbox{G}\!\to\! \mbox{SM}\! +\!\! \mbox{SM}$ & $+$ & $+$ & $-$ &  $-$  & $+$ \\   \hline
        inv. mode, $\mbox{G}\! \to \! \mbox{DM}\!\! + \!\!\mbox{DM} $ & $+$ & $+$ & $+$ &  $+$  & $-$ \\  
        \hline
        \hline
    \end{tabular} 
    \caption{ The benchmark parameters for the  spin-2 mediator production cross section $l^\pm N \to l^\pm N + \mbox{G} $ at the lepton fixed-target experiments.  Note that the $E_{\rm G}^{\rm cut}=x_{\rm cut} E_l$ is a typical minimum missing energy threshold 
    that is associated with the specific fixed-target facility. 
    \label{tab:BenchLeptonFTexp}
	}
\end{table}

\begin{figure*}[ht!]
	\center{\includegraphics[scale=0.12]{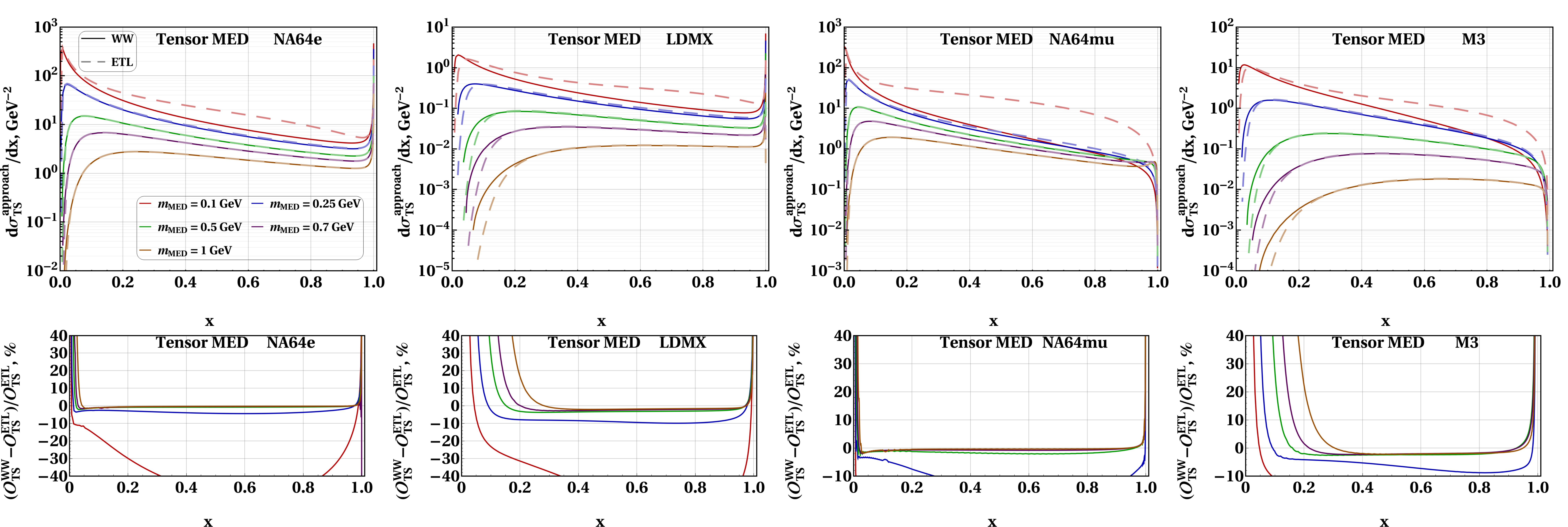}}
	\caption{ 
    The differential cross sections as a function of energy fraction $x$ in the case of tensor mediator for different approximations and the corresponding relative difference, where the columns correspond to NA64e, LDMX, NA64$\mu$, and M$^3$ experiments, respectively. 
    We also set~$\theta_{\rm max}~=~0.1$ and~$c_{ll}^{\rm G} = 1\;\mbox{GeV}^{-1}$.
    The curve colors correspond to fixed values of the mediator masses. 
    In the upper row we show the differential cross sections for different masses, where the solid and dashed lines correspond to the WW and ETL approximations, respectively.
    In  the bottom row the relative difference of differential cross sections between the WW and ETL approximations are shown.  
    }
	\label{fig:dsdxWWmedGravitonAllFFandLabs}
\end{figure*}

\subsection{The experiments to search for the invisible decays of MED}\label{sec:FixedTarExpParamInVisMode}

The mediator of DM can be produced in bremsstrahlung-like process where a lepton beam with the initial energy~$E_l$ incidents on a fixed target. 
In particular, the fraction~$x~=~E_{\rm miss}/E_l$ of the primary beam energy~$E_l$ can be  
carried  away in dark sector that leads to a missing energy deposition in a detector system. That 
can imply the invisible decay of spin-2 mediator into DM particles,  with $ 
\mbox{Br}(\mbox{G} \to \mbox{DM}+\mbox{DM})\simeq 1$. 
Moreover, the fixed-target experiments can probe the parameter space of light DM due to a high-energy particle beam and its relative large intensity.
In Tab.~\ref{tab:BenchLeptonFTexp} the crucial 
parameters of the considered fixed-target  experiments are shown.
\textit{NA64$e$}: the experiment is located in the north area of the European 
Organization for Nuclear Research~(CERN) on the H4 line of the super proton 
synchrotron~(SPS)~\cite{NA64:2017vtt}. 
\textit{NA64$\mu$}:  this experiment corresponds to the muon beam mode at the M2 line 
of the SPS in the north area of CERN, which uses the detector system structure of 
the NA64$e$ experiment~\cite{NA64:2024klw}.
\textit{LDMX}: the Light Dark Matter Experiment is a planned experiment with an electron beam
at Stanford Linear Accelerator Center~(SLAC) in which missing energy signatures are 
complemented by a unique technique to measure the missing momentum of 
electrons~\cite{Mans:2017vej,LDMX:2018cma}.
\textit{M$^3$}: this detector is considered  as complementary part of the LDMX, it has 
a similar detector base~\cite{Akesson:2022vza} and involves a muon beam. 


\section{Bremsstrahlung-like production of the tensor mediator}~\label{sec:BremsstrahlungLike}

The radiation of a dark matter mediator by a lepton scattering off a heavy nucleus can be 
represented as a $2~\to~3$ process:
\begin{equation}\label{eq:process2to3}
l^{\pm}(p)~+~N(P_i)~\rightarrow~l^{\pm}(p')~+~N(P_f)~+~\text{G}(k),
\end{equation}
where~$p~=~(E_{l},\vect{p})$, $p'~=~(E'_{l},\vect{p'})$ are 4-momenta of the incoming and outgoing leptons, respectively, $k~=~(E_{G},\vect{k})$ is 4-momentum of the dark matter mediator, $P_i~=~(M,0)$~and~$P_f~=~(P^0_f,\vect{P}_f)$ are \text{4-momenta} of the initial and final nucleus, respectively, $q~=~(q_0,\vect{q})~\equiv~P_i~-~P_f$ is 4-momentum transferred to the nucleus, and virtuality of the photon is~$
	t~\equiv~-q^2~=~-(P_i~-~P_f)^2
$.
The Mandelstam-like variables for the process~\eqref{eq:process2to3} can be introduced as:
\begin{equation*}
    \tilde{s} = (p' + k)^2 - m_l^2,
\quad
    \tilde{u} = (p - k)^2 - m_l^2,
\end{equation*}
\begin{equation}\label{eq:MandelstamDef2to3}
    \tilde{t} = (p - p')^2 - m_{\rm G}^2.
\end{equation}

In general, the interaction of an electromagnetic field~$A_{\mu}$ and a hadron can be effectively represented as~\cite{Schwartz:2014sze}:
\begin{equation}
	\mathcal{L}^{\rm nucl}_{\rm eff}  \supset - e A_{\mu} \mathcal{J}^{\mu},
\end{equation}
where ~$\mathcal{J}^{\mu}$ is a  hadronic current~\cite{Drell:1963ej,Berestetskii:1982qgu}. 
For heavy nucleus one can exploit a spin-0  form-factor with a good 
accuracy~\cite{Beranek:2013yqa}. 
Thus, the hadronic current in the momentum space reads as follows~\cite{Perdrisat:2006hj,Beranek:2013yqa}:
\begin{equation}
	\mathcal{J}^{\mu} = F(t)(P_f + P_i)^{\mu}.
\end{equation}
The nuclear form-factor in the laboratory frame is associated with charge density of nucleus through the Fourier transformation~\cite{Bjorken:2009mm,Kirpichnikov:2021jev,Tsai:1973py}.  
The elastic atomic Tsai-Schiff's~\cite{Tsai:1973py,Schiff:1953yzz} form-factor takes the following form:
\begin{equation}\label{TsaiFFdefinition11}
    F_{\rm TS}(t)  = 
    \frac{t}{t_{\rm a}~+~t}~\frac{t_{\rm d}}{t_{\rm d} + t},
\end{equation}
where $\sqrt{t_{\rm d}}~=~0.164 A^{-1/3} \text{GeV}$ is the typical momentum associated with the effective nuclear radius, $R_{\rm n}$, such that 
$t_{\rm d} =1/R_{\rm n}^2 $ and  
$\sqrt{t_{\rm a}}~=~1/R_{\rm a}$ is  typical momentum associated with the atomic radius, 
$R_{\rm a} = 111 Z^{-1/3}/m_e$.
Also, we take into account that~$|\vect{q}|~\lesssim~\mathcal{O}(100)~\mbox{MeV}$ and~$M~\propto~\mathcal{O}(100)~\mbox {GeV}$ imply the relation~$|\vect{q}|/M~\ll~1$.

The matrix element for the process~\eqref{eq:process2to3} can be written as:
\begin{equation}
	i \mathcal{M}_{2\to3}^{\rm G} 
= 
	i c^{\rm G}_{ll} e^2 \mathcal{L}^{\mu} \left(\frac{-i\eta_{\mu \nu}}{q^2}\right) \mathcal{J}^{\nu}
=
	i C_{\mathcal{M}} \mathcal{L}^{\mu} P_{\mu},
\end{equation}
where we denote $C_{\mathcal{M}} = c^{\rm G}_{ll} e^2 F_{\rm s}(q^2)/q^2$ and $P^{\mu}~=~(P_f~+~P_i)^{\mu}$. 
The corresponding lepton current $\mathcal{L}^{\mu} = \sum_{i=1}^{4} 
({\mathcal{L}_i})^{\mu}$ with mediator radiation consists of the terms:
\begin{align}
	  (\mathcal{L}_{1})_{\lambda} 
= &
	- i
	{\varepsilon^{*}}^{\mu \nu}(k)
 \nn \\ & \cdot
	\overline{u}(p') T_{\mu \nu }^{({ \rm G} l l)}(p' + k, p') \frac{\slashed{p}'+\slashed{k} + m_l}{\tilde{s}} \gamma_{\lambda}
	u(p), 
	\\
	  (\mathcal{L}_{2})_{\lambda}
= &
	-i
	{\varepsilon^{*}}^{\mu \nu}(k)
 \nn \\ & \cdot
	\overline{u}(p')
	\gamma_{\lambda} 
	\frac{\slashed{p} - \slashed{k} + m_l}{\tilde{u}}
	T_{\mu \nu }^{({ \rm G} l l)}(p, p - k)
	u(p),
	\\
	  (\mathcal{L}_{3})_{\lambda}
= &
	- i
	{\varepsilon^{*}}^{\mu \nu}(k)
	\overline{u}(p') \left( T_{\mu \nu \lambda}^{({ \rm G} l l \gamma)} / e  \right)
	u(p),
	\\
	  (\mathcal{L}_{4})_{\lambda}
= & 
	  + i
	{\varepsilon^{*}}^{\mu \nu}(k)
 \nn \\ & \cdot
	T_{\mu \nu \beta \lambda }^{({\rm G} \gamma \gamma )}(p - p', q )
	\overline{u}(p')
	\frac{\gamma^{\beta}}{\tilde{t} + m_{\rm G}^2}
	u(p).
\end{align}
The  the matrix element squared for the process~$2\to3$ can be written as:
\begin{equation}
	\overline{\left| \mathcal{M}_{2\to3}^{\rm G} \right|^2}
	=
	C_{\mathcal{M}}^2 
	\left|\mathcal{A}_{2\to3}^{\rm G} \right|^2,
\end{equation}
where the expression for the effective mediator production amplitude squared~$\left|\mathcal{A}_{2\to3}^{\rm G} \right|^2$ is  shown in appendix~\ref{sec:MatEl2to3Graviton}.

The ETL double differential cross section   for the bremsstrahlung-like production of mediator takes the following form~\cite{Liu:2016mqv,Liu:2017htz}:
\begin{multline}\label{eq:dsETL}
	\frac{d \sigma_{2\to3}}{d x d\cos(\theta_{\rm G})}
=
	\frac{(c^{\rm G}_{ll})^2 \alpha^3 Z^2}{4 \pi}
	\frac{|\vect{k}| E_p}{|\vect{P}| |\vect{k} - \vect{p}|}
\cdot \\ \cdot
	\int\limits_{t_{\rm min}}^{t_{\rm max}}
	dt
	\frac{F^2(t)}{t^2}
	\frac{1}{8 M^2}
	\int\limits_{0}^{ 2 \pi}
	\frac{d \phi}{2 \pi}
	\left|\mathcal{A}_{2\to3}^{\rm G} \right|^2,
\end{multline}
where~$\alpha~=~e^2/(4\pi)~\simeq~1/137$ is the fine structure constant,
 $x~=~E_{\rm G}/E_{l}$ is the fraction of the total energy of the mediator with respect to the energy of 
 original lepton,  $t_{\rm max}~=~t(Q_+)$ and~$t_{\rm min}~=~t(Q_-)$ are the maximum and minimum squares 
 of the 4-momentum and are introduced notations:
\begin{equation*}\label{eq:LimitstVirtualBremssETL}
	Q_{\pm} 
=
    \left|
	\frac{		|\vect{V}| \left[\tilde{u} + 2 M (M~-~V_0) \right] 
			\pm (M~-~V_0) \sqrt{ D_0  } 
		 }
		 {2 (M~-~V_0)^2 - 2 |\vect{V}|^2 }
   \right|,
\end{equation*}
\begin{equation*}
    V~\equiv~k-p,
\quad
    D_0~=~4 M^2 |\vect{V}|^2 + \tilde{u}^2  + 4 M d_{V_0} \tilde{u}.
\end{equation*}

The double differential cross section in the WW approximation takes the following form~\cite{Kim:1973he,Tsai:1973py}:
\begin{align}
&     \left.\frac{d \sigma ( p + P_i \rightarrow  p' + P_f + k )}{d(pk)d(kP_{i})}\right|_{WW} = 
\nn  \\   
&    
=    \frac{\alpha \chi}{\pi (p'P_i)}
    \left. \frac{d \sigma ( p + q \rightarrow  k+ p' )}{d(pk)} \right|_{ t = t_{\rm min} }.
    \label{DoubleDiffWW1}  
\end{align}
where the differential cross section for the Compton-like 
process
\begin{equation}
l^{\pm}(p)~+~\gamma(q)~\rightarrow~l^{\pm}(p')~+~\mbox{G}(k),
\end{equation}
can be written as:
\begin{equation}\label{8pissMM}
    \frac{d \sigma( p + q \rightarrow  k + p' )}{d (p k)}
=
    \frac{  (c^{\rm G}_{ll})^2 e^2
            \left|\mathcal{A}_{2\to2}^{\rm G} \right|^2}
        {8 \pi (s_2 - m_l^2)^2} ,
\end{equation}
where $\left|\mathcal{A}_{2\to2}^{\rm G} \right|^2$ is  the amplitude 
squared  of the  Compton-like processes~\cite{Voronchikhin:2022rwc}.

The flux of virtual photon~$\chi$ from nucleus in Eq.~(\ref{DoubleDiffWW1}) is expressed through the elastic form-factor~$F(t)$ as follows: 
\begin{equation}
\chi = 
   Z^2 \int\limits^{t_{\rm max}}_{t_{\rm min}} \frac{t - t_{\rm min}}{t^2} F^2(t) dt.
   \label{ChiDefininition1}
\end{equation}
The minimum value of virtuality in WW approximation~$t_{\rm min}^{\rm WW}$ reads:
\begin{equation} \label{tminDefinition1}
    t_{\rm min}^{\rm WW} \simeq U^2/(4E_l^2 (1-x)^2),
\end{equation}
where  we  denote the following function:
\begin{equation} 
\label{eq_vect_U}
    U \! \equiv \! m_l^2 \! - \! u_2 \! \simeq \!
    E_l^2 \theta_{\rm G}^2 x  +  m_{\rm G}^2 (1 \! - \! x)/x  +  m_l^2 x
\!>\! 
    0.
\end{equation}
In the so-called improved WW (IWW) approach~\cite{Liu:2017htz,Gninenko:2017yus} the dependence of $t_{\rm min}$ on $x$ and $\theta_{\rm G}$ in the flux derivation is omitted to simplify calculations, which is important for the integration that exploits, for instance, Monte-Carlo (MC)  methods~\cite{Kirpichnikov:2021jev}, such that 
\[
    t^{\rm IWW}_{\rm min} \simeq m_{\rm G}^4/(4 E_{l}^2).
\] 
Using Jacobian of the transformation from~$(k,p)$ and~$(k,\mathcal{P}_i)$ to~$\cos(\theta_{\rm G})$ and~$x~=~E_{\rm G}/E_{l}$ variables in the case of a ultra-relativistic incident lepton  in laboratory frame, one can get:
\begin{multline}\label{eq:dsWW}
	\left.\frac{d \sigma ( p + P_i \rightarrow  p' + P_f + k )}{dx d\cos(\theta_{\rm G})}\right|_{\rm WW}
	=  
	\frac{\alpha \chi}{ \pi }
\cdot \\ \cdot
	\frac{E_l^2 x \beta_{\rm G}}{1-x}
	\left. \frac{d \sigma ( p + q \rightarrow  k + p' )}{d(pk)} \right|_{ t = t_{\rm min} },
\end{multline}
where~$\beta_{\rm G}~=~\sqrt{1 - m_{\rm G}^2/(x E_l)^2}$ is the typical velocity of mediator.

\section{The bremsstrahlung-like cross sections }\label{sec:crossSectionsBremssLike}

In this section, we compare the cross sections in the  WW and ETL approaches 
for fixed-target experiments. 
\begin{figure}[ht!]
	\center{\includegraphics[scale=0.55]{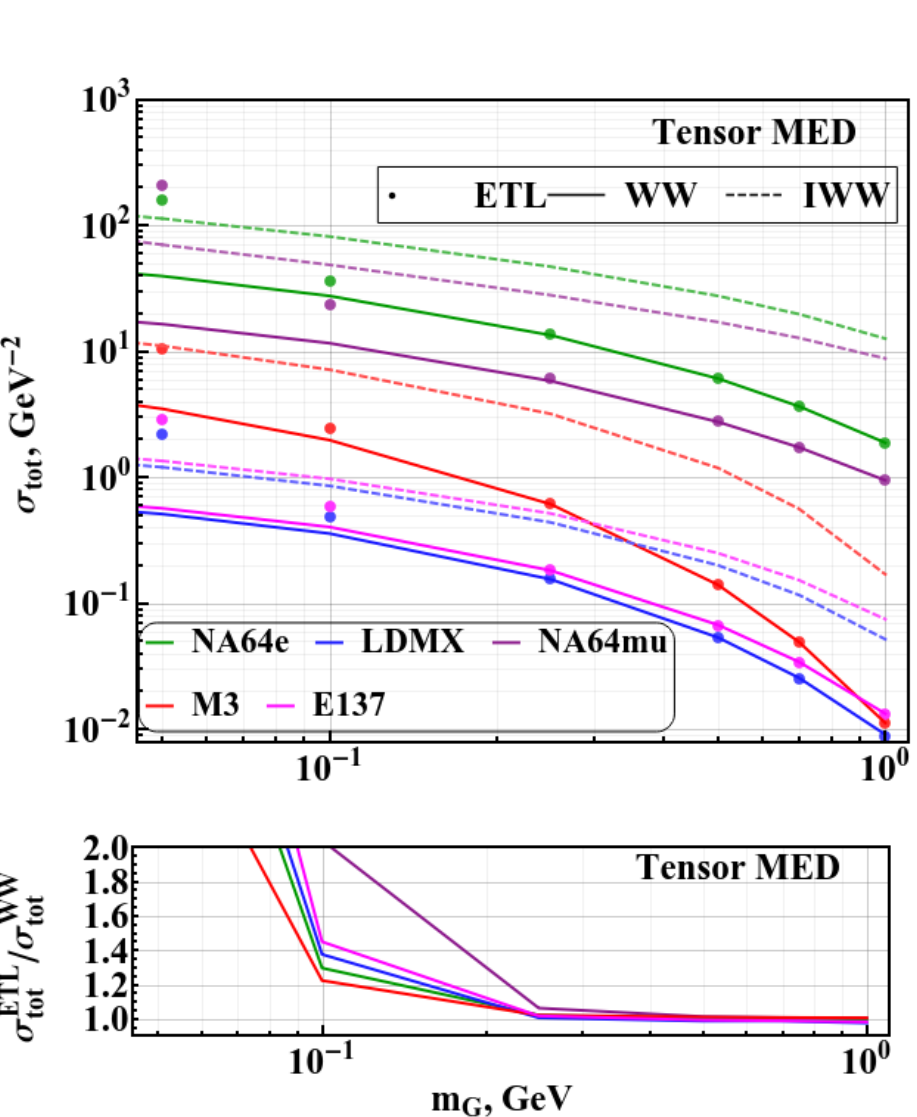}}
	\caption{ 
    The total cross sections~$\sigma_{\rm tot}$ for tensor mediator production with typical angle cut~$\theta_{\rm max}~=~0.1$, where the benchmark  Tsai-Schiff's form-factor is chosen. 
    The green, blue, purple, red and magenta lines correspond to NA64e, LDMX, NA64$\mu$, M$^3$, and E137 experiments, respectively. 
    Solid line, dashed line and dots correspond to calculations for WW, IWW, and ETL approximations, respectively.
    }
	\label{fig:dsWWmedGravitonAndRelDiffFF}
\end{figure}
The total and differential cross sections are calculated by the numerical integration of the 
amplitude squared over the phase space for  the specific experimental setup.
Moreover, Tsai-Shiff's form-factor (\ref{TsaiFFdefinition11}) is used as a benchmark one  
for the cross section in the WW and ETL  
approach~\cite{Bjorken:2009mm,Chen:2017awl,Kahn:2018cqs,Kirpichnikov:2021jev} throughout 
the paper.

We note that the impact of different form-factors parametrization on cross section was studied in 
Ref.~\cite{Voronchikhin:2022rwc,Voronchikhin:2024vfu}.
An optimal value of the maximum angle of $\mbox{G}$-emission is chosen to be $\theta_{\rm max} \simeq 0.1$, 
since the dominant contribution to the cross section occurs in the peak-forward 
region~\cite{Bondi:2021nfp,Oberhauser:2024ozf,Voronchikhin:2024vfu}.

\subsection{The differential cross section for invisible mode experiments
\label{SecDiffCS11}}

In this section, we discuss only experiments that aim to the search  for the
invisible mode decay of tensor mediator (see e.~g.~Tab.~\ref{tab:BenchLeptonFTexp}). 
In Fig.~\ref{fig:dsdxWWmedGravitonAllFFandLabs} we show   WW and ETL differential cross sections for the specific parameter set and the comparison between the corresponding approaches.


To be more specific, the relative difference between the WW and ETL differential cross sections for 
various experiments  can be described as follows. 
\textit{NA64e}: For the relatively large masses $m_{\rm G} \gtrsim 100~\mbox{MeV}$,
the using of WW approximation leads to overestimation at the level of~$\mathcal{O}(10)~\%$ 
around~$x~\simeq 1$ and~$x~\ll  1$, the discrepancy is negligible, $\lesssim \mathcal{O}(1)~\%$, 
for intermediate energy fraction range~$x~\lesssim~1$.
However, for masses $m_{\rm G} \lesssim 100~\mbox{MeV}$,
WW approximation leads to underestimation at the level of~$\gtrsim \mathcal{O}(10)~\%$ for the intermediate values $x~\lesssim~1$.
\textit{NA64$\mu$}: 
The relative difference for that one is similar to the NA64e case, although  a sizable discrepancy, $\gtrsim \mathcal{O} (50)\%$, appears  for the masses~$m_{\rm G} \lesssim 200~\mbox{MeV}$. 
 \textit{LDMX} and  \textit{M$^3$}:
the behavior of the relative differences is similar to the shapes of NA64$e$ and NA64$\mu$, respectively, however the $\mathcal{O} (10)\%$ 
overestimation  for the heavy masses, $m_{\rm G}\gtrsim 100~\mbox{MeV}$,
occurs in the intermediate range $x\lesssim 1$.

It is important to emphasize that for a tensor mediator with masses~
$m_{\rm G}\gtrsim 100\;\mbox{MeV}$ one has a fairly well agreement between the differential cross sections calculated in the WW and the ETL, for all  experiments of interest.
However, a sizable  difference, $\gtrsim \mathcal{O}(50)~\%$, between corresponding approaches arises due to the rapid increasing of the differential ETL cross section for small mass range~$m_{\rm G}~\lesssim~100\;\mbox{GeV}$.
Indeed, the $ l N \to l N \rm{G} $ amplitude squared  in the ETL method (see 
e.~g.~Eq.~(\ref{MainAsquared2to3})) includes the typical terms of negative 
mass power, $\propto 1/m_{\rm G}^2$ and $\propto 1/m_{\rm G}^4$, that arise
due to the summation over longitudinal polarizations (\ref{PolSumEq}) of the
light spin-2 mediator.   For the $l \gamma^* \to l {\rm G}$ amplitude squared in  
the  WW the regarding mass terms are absent, see 
Ref.~\cite{Voronchikhin:2022rwc,Gill:2023kyz}. As a result, in the ETL cross section
for relatively large masses $m_{\rm G}\gtrsim 100~\mbox{MeV}$,
the typical terms $\propto 1/m_{\rm G}^2$ and $\propto 1/m_{\rm G}^4$ can be negligible 
with the respect to the terms of $\propto m^0_{\rm G}$, $\propto m^2_{\rm G}$, 
$\propto m^4_{\rm G}$,  etc. Such that, the differential cross sections for
both the WW and the ETL methods are in a good agreement, implying the the intermediate 
energy fraction range, $ x \lesssim 1$.  

We note  that the typical 
shape of the relative difference between the WW and ETL differential cross 
sections  for tensor-mediator masses~$ m_{\rm G} \gtrsim 100~\mbox{MeV}$ is similar
to the one calculated for the spin-0 and spin-1 mediators~(see 
e.~g.~Ref.~\cite{Voronchikhin:2024vfu} for detail).

To conclude this sub-seсtion we note  that the differential cross section and the corresponding 
relative difference in ETL and WW calculations are similar for both the E137 and the LDMX 
experiments.  These experiments have a similar target and beam energy parameters,
which are given in the Tab.~\ref{sec:FixedTarExpParamInVisMode}. 

\subsection{The total cross section}

In Fig.~\ref{fig:dsWWmedGravitonAndRelDiffFF} we show the total cross section as a function of $m_{\rm G}$ and the regarding ration between the WW and ETL  approach for benchmark
fixed-target experiments presented in Tab.~\ref{tab:BenchLeptonFTexp}.  
That implies the typical benchmark angle cut~$\theta_{\rm max}~=~0.1$.  
For relatively large masses $m_{\rm G} \gtrsim 200~\mbox{MeV}$, the relative difference between two 
approaches is at the level of~$\mathcal{O}(1)~\%$, however relative difference can reach~$\mathcal{O}(20)~\%$ 
for the tensor-mediator mass range below~$m_{\rm G} \lesssim 100~\mbox{MeV}$. 

The origin of discrepancy in the small mass region is discussed  in Sec.~\ref{SecDiffCS11}, i.~e.~ it can also be referred  to the   typical $\propto 1/m^2_{\rm G}$
and $\propto 1/m_{\rm G}^4$ terms that arise in the ETL amplitude squared for the process 
$l N \to l N {\rm G}$. 

We address the typical range of the large masses  $m_{\rm G}~\gtrsim~200~\mbox{MeV}$  as  a region of 
robust calculation, for which WW and ETL approaches yield fairly good 
agreement. That can be crucial for the fixed-target experiments~\cite{Voronchikhin:2022rwc,Voronchikhin:2023znz}
that search for the 
bremsstralung-like production of the tensor mediator, $l N \to l N {\rm G}$,
followed by its invisible decay 
into  pair of DM particles, $\mbox{G}\to \mbox{DM} + \mbox{DM}$.
The regarding sensitivities to the invisible mode of $\mbox{G}$ can be found 
elsewhere Refs.~\cite{Voronchikhin:2022rwc,Voronchikhin:2023znz}, that implies WW approach for the 
$\mbox{G}$-strahlung cross section calculation.  

In the next section, we focus on the visible mode decay of the light spin-2  mediator 
into pair of  SM particles, $\G \to\gamma \gamma (e^+e^-)$, 
that exploits the results of the present section.

\section{Visible mode}\label{sec:VisModConstr}

\begin{figure*}[ht!]
	\center{\includegraphics[scale=0.475]{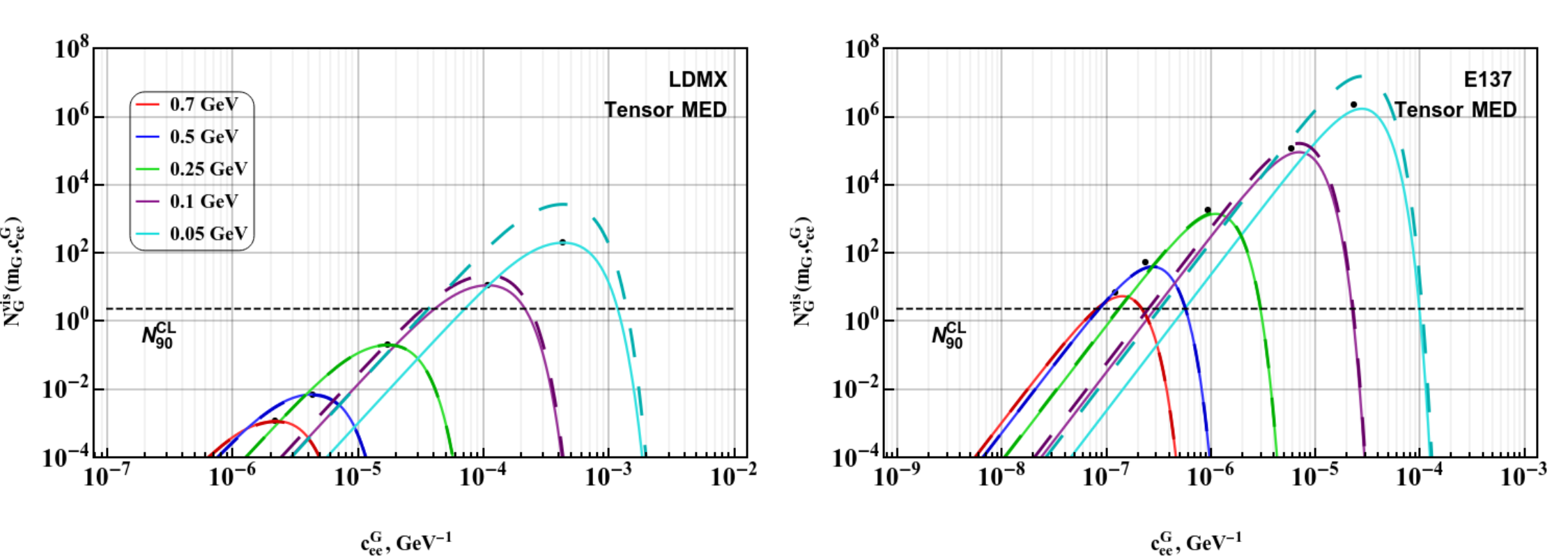}}
	\caption{ Number of signal events~\eqref{eq:NvisMod} as a function of the coupling constant $c^\G_{ee}$ for the typical set of masses. Left and right panels correspond to the LDMX ($\mbox{EOT}=10^{15}$) and E137 ($\mbox{EOT}=1.87\times 10^{20}$).   
    The number of signal events for the ETL~\eqref{eq:dsETL} and WW~\eqref{eq:dsWW} methods corresponds to the dashed and solid lines, respectively. 
    }
	\label{fig:Nvis}
\end{figure*}

If the  dark matter is heavier than its mediator, $m_\G~\lesssim~2~m_\DM$, 
then the corresponding mediator can only 
decay into SM particles due to the kinematical conditions.
A signal in visible mode means that the mediator passes through the target and 
decays before  the detector system. 
The corresponding thermal curves for the tensor mediator can be 
found in the Ref.~\cite{Kang:2020huh}, implying that 
the DM consists of scalar, vector, or Dirac particles. 
Remarkably, the E137 experiment~\cite{Bjorken:1988as} can yield a new constraint for the tensor mediator due to the large collected statistics, where  the spin-2 mediator
visibly decays into ${\rm G }~\to~\gamma\gamma$ and $\G~\to~e^+e^-$ in the 
 mass range~$m_{\rm G}~\gtrsim~\mathcal{O}(100)~\mbox{MeV}$. In what follows, we 
 consider the  spin-2 mediator, which predominantly couples to 
 photons and electrons~\cite{Kang:2020huh}.

\subsection{The experiments to search for the visible decays of the tensor mediator}

\textit{E137}: The experiment E137 at Stanford Linear Accelerator Center (SLAC) was originally 
designed to search for axion-like particles as neutral metastable 
particles~\cite{Bjorken:1988as}. 
 The null-result of E137 experiment can be exploited to set constraints on parameter space 
of spin-2 boson.
Particles of hidden sector can be produced by a 20 GeV electron beam incident on a fixed target 
that consists of a set of aluminum plates interlaced with cooling water.
The mediator produced by $\G$-strahlung reaction  can pass 179 m of shielding  and 
decay on $204~\mbox{m}$ long open region before a detector. 
In particular, the detector of E137 experiment consists of an 8-radiation length 
electromagnetic shower calorimeter that can detect charged particles or photons produced by the 
mediator.
For the E137 experiment the electrons accumulated on target (EOT) are set to be at the level of~$\mbox{EOT} = 1.87\times10^{20}$. 

\textit{LDMX}: We also discuss for completeness the LDMX  expected reach for  the visible mode, the corresponding 
experimental parameters are provided in Section~\ref{sec:FixedTarExpParamInVisMode}. We set the expected statistics of the LDMX to be $\mbox{EOT}~=~10^{15}$.

\subsection{Signal events for visible mode}

The cut for the mediator energy ratio~$x_{\rm cut}~=~E^{\rm cut}_{\rm G}/E_l $ is used in order to  specify  the typical missing energy signature of the considered fixed-target experiments (see e.~g.~Tab.~\ref{tab:BenchLeptonFTexp}).
Note that  the total cross section of bremsstrahlung-like tensor-mediator production for the specific experiment  is:
    $$\sigma_{\rm tot} (E_e) = \int\limits^1_{x_{\rm cut}} \! \! dx \! \! \int\limits^{\theta_{\rm max}}_0 \!\! 
    d \theta_{\rm G} 
    \frac{d\sigma_{2\to 3}}{dx d\theta_{\rm G}}(E_e),$$
where the double differential cross section is shown in Sec.~\ref{sec:BremsstrahlungLike}  for various approaches, $\theta_{\rm G}$ is an angle between initial beam direction and momentum of the produced~$G$.

 Assuming production of a mediator in the first  radiation length, one can estimate the number of 
 the  bremsstrahlung-like mediator production for fixed-target facilities as 
 follows~\cite{NA64:2022rme}: 
\begin{equation}\label{eq:Nbremss}
N^{\rm brem. }_{\rm G} \simeq \mbox{EOT}\cdot \frac{\rho N_{\rm A}}{A} L_{\rm T} \cdot \sigma_{\rm tot}(E_e),
\end{equation}
where $X_0~=~8.9\;\mbox{\rm cm}$ is a radiation length of the aluminium,
$L_{\rm T}^{\rm LDMX}~\simeq~0.4X_0$, $L_{\rm T}^{\rm E137}~\simeq~X_0$ are effective radiation lengths of the LDMX and E137 experiments, respectively.  
The decay length of the new particle in the lab frame is:
\begin{equation}\label{eq:LenDecay}
    l_{\rm G} = \frac{E_{\rm G}}{m_{\rm G}} \frac{1}{\Gamma^{\rm tot}_{\rm G}},
\end{equation}
where~$\Gamma^{\rm tot}_{\rm G}$ is a total decay width $\Gamma^{\rm tot}_{\rm G}~=~\Gamma_{G \to ee}~+~\Gamma_{G \to \gamma \gamma}$
with $\Gamma_{G \to ee}$  and $\Gamma_{G \to \gamma \gamma}$
being a partial  decay widths to the electron and the photon, 
respectively~\cite{Lee:2013bua}:
\begin{equation}
         \Gamma_{G \to ee} \label{Gto2psiDecayWidth} 
 =
    \frac{(c_{ee}^{\rm G})^2 m_{\rm G}^3 }{160 \pi} 
       \left(\! 1 + \frac{8}{3}\frac{m_e^2}{m_{\rm G}^2} \! \right)
    \left(\! 1 - 4 \frac{m_e^2}{m_{\rm G}^2}\! \right)^{3/2},    
\end{equation}
\begin{equation}
         \Gamma_{G \to \gamma \gamma} \label{Gto2VDecayWidth}
 =
    13 \frac{(c_{\rm \gamma \gamma}^{\rm G})^2 m_{\rm G}^3}{960 \pi}.    
\end{equation}
 The coupling constants are chosen  to be~$c_{\gamma\gamma}^{\rm G}~=~c_{ee}^{\rm G}$.
 For the  decay length into photons, one can get:
\begin{multline*}\label{eq:LenDecayApprox}
    l_{{\rm G} \to \gamma \gamma}
\simeq
    4.5 \cdot 10^{5}~\mbox{cm}
 \\ \times
    \left( \frac{E_{\rm G}}{10~\mbox{GeV}} \right)
    \left( \frac{0.1~\mbox{GeV}}{m_{\rm G}} \right)^{4}
    \left( \frac{10^{-6}~\mbox{GeV}^{-1}}{c_{\rm ee}^{\rm G}} \right)^{2}.
\end{multline*}
Next, using the thick target approximation, the number of signal event in visible mode takes the following form~\cite{Berlin:2018bsc}:
\begin{equation}\label{eq:NvisMod}
    N^{\rm vis. }_{\rm G} \simeq N^{\rm brem. }_{\rm G} \times (e^{-L_{\rm sh}/ l_{\rm G}} - e^{-L_{\rm tot}/ l_{\rm G}}),
\end{equation}
where~$L_{\rm sh}$~is the length of 
shield, $L_{\rm tot}~=~L_{\rm sh}~+~L_{\rm dec}$ with $L_{\rm dec}$~being the length of the fiducial volume.
In Eq.~(\ref{eq:NvisMod}) we suppose that 
the dominant part of the electron energy transfers to the tensor mediator, such that $E_e~\simeq~E_\G$.
The signals from secondary positrons production in the target~\cite{Marsicano:2018krp}  and the 
absorption  processes~\cite{Liu:2017htz} are neglected in our calculation. 
The shielding parameters for LDMX~\cite{Berlin:2018bsc} and E137~\cite{Andreas:2012mt,Liu:2017htz} read, respectively:
\begin{equation}
    L_{\rm sh}^{\rm LDMX}~=~43\;\mbox{cm}, 
\quad
    L_{\rm tot}^{\rm LDMX}~=~315\;\mbox{cm},
\end{equation}
\begin{equation}
    L_{\rm sh}^{\rm E137}~=~179\;\mbox{m},
\quad
    L_{\rm dec}^{\rm E137}~=~204\;\mbox{m}.
\end{equation}

In Fig.~\ref{fig:Nvis} we show the number of signal events for LDMX and E137 
experiments.

\begin{figure}[tbh]
	\center{\includegraphics[scale=0.35]{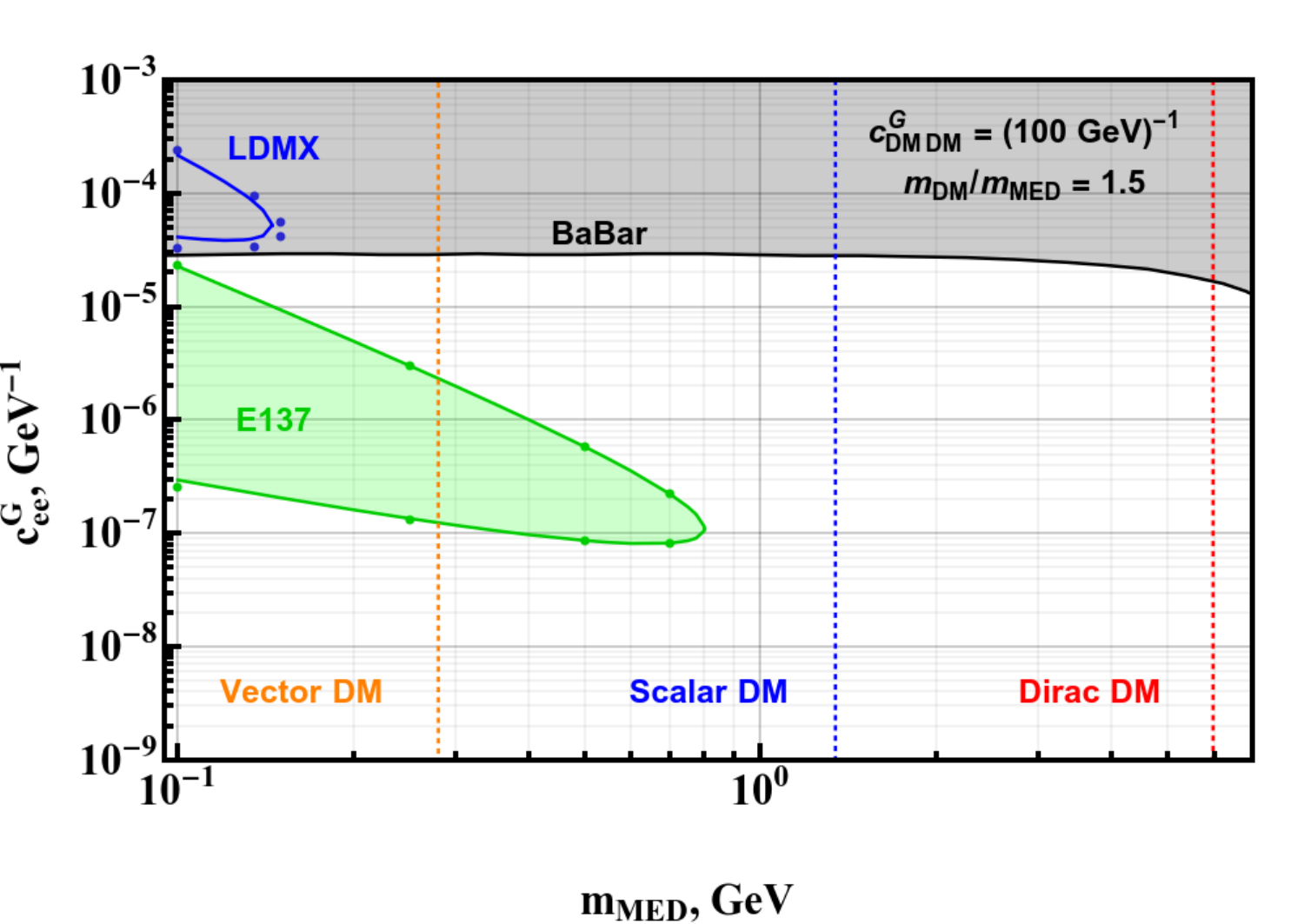}}
	\caption{ The experimental reach at $90\, \%$ C.L. as the function of MED mass for the 
    LDMX ($\mbox{EOT}= 10^{15}$) 
    and E137 ($\mbox{EOT}=1.87\times 10^{20}$) fixed-target facilities in the 
    case of the visible mode. Green shaded region is the current reach of the  electron beam 
    dump      E137~\cite{Bjorken:1988as} experiment, and the blue area corresponds to  the 
    planed  sensitivity of the LDMX experiment.
    Gray shaded region is the current constraint of BaBar 
    experiment~\cite{Kang:2020huh}.
    The relic DM density is shown by the orange, red and blue dashed lines for vector, fermion 
    and scalar DM,  respectively~\cite{Kang:2020huh}, also all these lines imply~$c_{\rm 
    DM\;DM}^{\rm G}~\simeq~(100~\mbox{GeV})^{-1}$ and ~$m_{\DM}~=~1.5~m_{\G}$.  
    The constraints on the coupling constant using ETL in the case of E137 and LDMX experiments  
    calculations are shown by green and blue dots, respectively.
    }
	\label{fig:1}
\end{figure}

\subsection{The experimental limits}
Assuming background free case and zero signal events observed at fixed-target experiments 
we require  $90~\% \mbox{C. L.}$ upper limit on the number of spin-2 mediator decays to be 
$N^{\rm CL}_{90\%}\simeq 2.3$ according to Poisson statistics.  For each mediator mass 
$m_{\G}$, the range of coupling constrained 
$c^{\rm low}_{ee}(m_\G) \lesssim  c^\G_{ee} \lesssim  c_{ee}^{ \rm up}(m_\G)$
is defined by inequality $N^{\rm vis. }_\G \gtrsim N^{\rm CL }_{90\%}$, see Fig.~\ref{fig:Nvis}.
The values above $c_{ee}^{ \rm up}(m_\G)$ correspond to short-lived  decaying  before the shield 
escaping.
The values below $c^{\rm low}_{ee}$ correspond to too small signal yield of $\G$ decaying beyond 
the fiducial volume.

In Fig.~\ref{fig:1} we show the $90\%$ C.L. sensitivity region of both LDMX and 
E137 experiments in the $\G$-parameters space for the background free case.  
These results demonstrate  that the E137 experiment has been ruled out the 
the parameter space of spin-2 DM mediator at the level of 
$8\times 10^{-8}\, \mbox{GeV}^{-1} \lesssim  c^\G_{ee} \lesssim 10^{-5} \mbox{GeV}^{-1}$
for the typical  masses in the range   $100 \mbox{MeV} \lesssim m_\G \lesssim 800~\mbox{MeV}$, 
that corresponds to the $\mbox{EOT}=1.87\times 10^{20}$. The bounds  of the LDMX 
experiment for  $\mbox{EOT}=10^{15}$ have been already ruled out by BaBar data on 
mono-photon, $e^+ e^- \to \G \gamma$, searches for the spin-2 DM mediator. 
The regarding lepton collider bound  applies to the typical parameter space of
the spin-2 mediator with $m_\G~\lesssim~7~\mbox{GeV}$ and 
$c^\G_{ee} \lesssim 3\times 10^{-5}~\mbox{GeV}$.

In addition we note that the thermal production mechanism of DM involving spin-2  
mediator have been analyzed in detail in Ref.~\cite{Kang:2020huh}. 
To be more specific, in Fig.~\ref{fig:1} we show the typical relic-abundance
 curves for the scalar, vector and Dirac DM  adapted from Ref.~\cite{Kang:2020huh}. 
 These lines imply the typical DM couplings 
 $c^\G_{SS}~=~c^\G_{\psi \psi}~=~c^\G_{VV}~=~10^{-2}~\mbox{GeV}^{-1}$ and 
 benchmark mass ratio $m_{\DM}/m_{\G}=1.5$, that leads to the visible decays of spin-2 mediator. 
 Remarkably,  the E137 experiment has been ruled out the vector DM  with 
 $m_{V}~\simeq~300~\mbox{MeV}$ for the coupling constant in the range 
 $10^{-7}~\mbox{GeV}^{-1}~\lesssim~c^\G_{ee}~\lesssim~3\times10^{-6}~\mbox{GeV}^{-1}$. 
 In addition, we note that E137 searches  are not sensitive to both scalar and Dirac DM.

\section{Conclusion\label{sec:Conclusion}}
In the present paper, we have discussed in detail the
calculation of the DM mediator production cross sections in the
case of ETL and WW approaches for fixed-target experiments, such as NA64e, LDMX, NA64$\mu$, 
and M$^3$.  Tensor massive particle $\G$ is chosen as hidden-sector mediator, that is produced in
the bremsstrahlung-like reaction $lN~\to~lN\G$ followed by its invisible decay into DM pair,
$\G~\to~\DM~+~\DM$.   Our study shows that for mass range $m_\G~\gtrsim~200~\mbox{MeV}$ the 
agreement between ETL and WW methods can be at the level of $\lesssim \mathcal{O}(1)~\%$ for all 
fixed-target experiments of interest.  

In addition, we have studied the properties of the spin-2 particle production 
in the bremsstrahlung-like reaction $ lN~\to~lN\G$ and its visible decays, 
$ \G~\to~\gamma\gamma$ and $\G~\to~e^+e^-$, for the LDMX and E137 experiments. 
We have calculated the expected sensitivity to spin-2 DM mediator of the LDMX 
experiment for the statistics $\mbox{EOT}=10^{15}$ and 
have shown that the regarding parameter space was ruled out by mono-photon BaBar 
searches, $e^+e^-~\to~\G\gamma$.  
We derived novel constraints of  the E137 experiment that are associated with the coupling constant region
$10^{-7}~\mbox{GeV}^{-1}~\lesssim~c^\G_{ee}~\lesssim~10^{-5}~\mbox{GeV}^{-1}$ 
for the mass range   $100~\mbox{MeV}~\lesssim~m_\G~\lesssim~800~\mbox{MeV}$, that corresponds to
the statistics of $\mbox{EOT}=1.87\times 10^{20}$ electrons  accumulated 
on target.

\section{Acknowledgements}
This work  is supported by  RSF  grant no. 24-72-10110.  
We would like to thank A.~Arbuzov, R.~Dusaev, D.~Gorbunov,  
S.~Gninenko, V.~Kim,
M.~Kirsanov, N.~Krasnikov, E.~Kriukova, V.~Lyubovitskij, S.~Ramazanov,
A.~Shevelev,  and  A.~Zhevlakov  for fruitful discussions.

\appendix

\section{Tensor mediator
\label{FPaction}}

We focus on the linearized theory of the massive spin-2 
particle~\cite{Hinterbichler:2011tt,deRham:2014zqa}:
\begin{multline}\label{eq:ActionFierzPauli}
	S_{\rm FP} 
=
	   \frac{1}{2}
	\int d^{D} x
	\left[
	\partial_{\sigma} G_{\mu \nu}  \partial^{\sigma} G^{\mu \nu}
	-   2 \partial_{\mu} G_{\nu \sigma}  \partial^{\nu} G^{\mu \sigma}
+ \right. 
\\ 
\left. +  
	2 \partial_{\mu} G^{\mu \nu}  \partial_{\nu} G
	-    \partial_{\sigma} G  \partial^{\sigma} G
	-    m_{\rm G}^2 (G_{\mu \nu} G^{\mu \nu} - G^2)
	\right],
\end{multline}
where~$\eta_{\mu \nu}$ is Minkowski metric.
Next, by varying the action~\eqref{eq:ActionFierzPauli} over the 
field~$G^{\mu \nu}$,  one can obtain the Fierz-~Pauli equation for a massive 
spin-2 particle that takes the following form~\cite{Fierz:1939ix}:
\begin{multline}\label{eq:MassSpin2eqMotionFreeAll}
	      \partial_{\sigma} \partial^{\sigma} G^{\mu \nu}
	-   \partial_{\sigma} \partial_{\mu} {G^{\sigma}}_{\nu}
	-   \partial_{\sigma} \partial_{\nu} {G^{\sigma}}_{\mu}
	+   \eta_{\mu \nu} \partial_{\lambda} \partial_{\sigma} G^{\lambda \sigma}
+  \\ + 
        \partial_{\mu} \partial_{\nu} G
	-   \eta_{\mu \nu} \partial_{\sigma} \partial^{\sigma} G
	+   m_{\rm G}^2 (G_{\mu \nu} - G)
	=   0.
\end{multline} 
After some simplifications, one can get the \mbox{Klein-Gordon} equation,~$(\partial_{\sigma}\partial^{\sigma}~+~m_{\rm G}^2)G_{\mu \nu}~=~0$, with additional conditions 
\begin{equation}
G^\mu_\mu \equiv G~=~0, \qquad\partial^{\mu} G_{\mu \nu}~=~0.
\label{TransvTrsllCond}
\end{equation}

By taking into account the symmetry of the field~$G_{\mu \nu}$, the number of freedom degrees 
is~$D(D+1)/2$, also the number of freedom degrees decreases to~$(D^2~-~D~-~2)/2$ 
due to Eq.~(\ref{TransvTrsllCond}).
In case of a spin-2 particle, the sum over polarizations~${\varepsilon}^{\mu \nu} (\vect{p},\lambda)$, takes the form:
\begin{multline}
	P^{\mu \nu \alpha \beta} 
= 
	\sum\limits_{\lambda}
	{\varepsilon}^{\mu \nu} (\vect{p},\lambda)
	{\varepsilon}^{\alpha \beta} (\vect{p},\lambda)
 \\ =
	\frac{1}{2}
	\left(
		P^{\mu \alpha } P^{\nu \beta } + P^{\mu \beta } P^{\nu \alpha }
	\right)
	-   \frac{1}{D - 1 - \xi}  P^{\mu \nu  } P^{ \alpha \beta },
    \label{PolSumEq}
\end{multline}
and the propagator of the tensor mediator reads as:
\begin{equation}
	{\mathcal{D}_{\rm G}}^{\mu \nu \alpha \beta}(p)
=
	i P^{\mu \nu \alpha \beta} / \left( p^2 - m_{\rm G}^2\right),
\end{equation}
where~$P^{\mu \nu }~=~-\eta^{\mu \nu }~+~(1~-~\xi)p^{\mu}p^{\nu}/m_{ \rm G}^2$ is the 
expressions for the sum over polarizations of a vector particle; $\xi~=~1$ is Feynman gauge 
(massless graviton) and~$\xi~=~0$ is Landau gauge (massive graviton).

\begin{widetext}

\section{Matrix element
\label{sec:MatEl2to3Graviton}}
The amplitude squared for the process $l N \to l N \G$ reads
\begin{equation}
    \left| \mathcal{A}_{2\to3}^{\rm G} \right|^2 
=
    \frac{ R^{(0)} + R^{(2)} m_l^2 + R^{(4)} m_l^4
         }{ 12 m_{\rm G}^4 \tilde{s}^2 \tilde{u}^2
            \left[  \tilde{s} + \tilde{u} + t - m_{\rm G}^2  \right]^2 
         },
         \label{MainAsquared2to3}
\end{equation}
where the typical terms are expressed through the sums  
\begin{align}
    R^{(0)} = \sum_{i = 0}^{6} m_{\rm G}^{2*i} R^{(0)}_{i},
\quad
    R^{(2)} = \sum_{i = 1}^{5} m_{\rm G}^{2*i} R^{(2)}_{i},
\quad
    R^{(4)} = \sum_{i = 2}^{4} m_{\rm G}^{2*i} R^{(4)}_{i},
\end{align}

\begin{align}
    R^{(0)}_{6} 
&&  = 
    3 \left\{
        \tilde{s}^2 \left[  P^2 t - 4 (p',P)^2  \right]
    -   2 \tilde{s} \tilde{u} \left[  P^2 t + 4 (p,P) (p',P)  \right]
    +   \tilde{u}^2 \left[  P^2 t - 4 (p,P)^2  \right]
   \right\},
\end{align}
\begin{align} %
     R^{(0)}_{5} 
& = 
        3 \tilde{s}^3 
        \left\{  P^2 (\tilde{u}-2 t)+8 (p',P)^2  \right\}
    +  \nonumber  \\  & + 
        6 \tilde{s}^2 
        \left\{
            P^2 
            \left[  -t^2 + 2 t \tilde{u} + 4 \tilde{u}^2  \right]
        +   8 (p,P) (p',P) \tilde{u}
        +   4 (p',P)^2 \left[  t + \tilde{u}  \right]
        \right\}
   + \nonumber  \\  & +   
        3 \tilde{s} \tilde{u} 
        \left\{
            14 P^2 t^2 
        +   4 P^2 t \tilde{u}
        +   P^2 \tilde{u}^2
        \right\}
    + \nonumber  \\  & +   
        3 \tilde{s} \tilde{u} 
        \left\{
        -   4 t \left[  (p,P)^2 - 12 (p,P) (p',P) + (p',P)^2  \right]
        +   8 (p,P) \tilde{u} \left[  (p,P) + 2 (p',P)  \right]
        \right\}
    + \nonumber  \\  & +
        6 \tilde{u}^2 \left\{  t + \tilde{u}  \right\} \left\{  4 (p,P)^2 - P^2 t  \right\},
\end{align}
\begin{align} %
    R^{(0)}_{4} 
& = 
    - 3 
    \left\{
        \tilde{s}^4 
        \left[  -P^2 t + 2 P^2 \tilde{u} + 4(p',P)^2  \right]
    + \right. \nonumber \\ & +    
        2 \tilde{s}^3 
        \left[
            P^2 
            \left[ -t^2 + 3 t \tilde{u} + 9 \tilde{u}^2 \right]
       +    4 (p,P) (p',P) \tilde{u}
       +    4 (p',P)^2 \left[  t + \tilde{u}  \right]
       \right]
    + \nonumber \\ & +    
        \tilde{s}^2 
        \left[ 
            P^2
            \left[  -t^3 + 30 t^2 \tilde{u} + 26 t \tilde{u}^2 + 18 \tilde{u}^3  \right]
        +   4 (p,P)^2 \tilde{u} \left[  \tilde{u} - 2 t  \right]
       \right]
    + \nonumber \\ & +    
        \tilde{s}^2 
        \left[ 
            16 (p,P) (p',P) \tilde{u} \left[  5 t + \tilde{u}  \right]
        +   4 (p',P)^2 \left[  t^2 + \tilde{u}^2  \right]
       \right]
    + \nonumber \\ & +
        2 \tilde{s} \tilde{u}
        \left[
            P^2 \left[  15 t^3 + 15 t^2 \tilde{u} + 3 t \tilde{u}^2 + \tilde{u}^3  \right]
       +    4 (p,P)^2 \left[  \tilde{u}^2 - 3 t^2  \right]
        \right]
    + \nonumber \\ & +
        2 \tilde{s} \tilde{u}
        \left[
            4 (p,P) (p',P) \left[  t + \tilde{u}  \right] \left[9 t + \tilde{u}  \right]
       -    4 (p',P)^2 t \left[ 3 t + \tilde{u} \right]
        \right]
    + \nonumber \\ & \left. +   
        \tilde{u}^2 \left[  t + \tilde{u}  \right]^2 \left[ 4 (p,P)^2 - P^2 t \right]
    \right\},
\end{align}
\begin{align} %
    R^{(0)}_{3} 
&  = 
    \tilde{s} \tilde{u} 
    \left\{
        3 \tilde{s}^4 P^2
    +   6 \tilde{s}^3 P^2 \left[ 2 t + 7 \tilde{u} \right]
    + \right. \nonumber \\ & +   
        3 \tilde{s}^2 
        \left[
            P^2 \left[  23 t^2 + 32 t \tilde{u} + 18 \tilde{u}^2  \right]
        -   4 (p,P)^2 t 
        +   40 (p,P) (p',P) t
        -   4 (p',P)^2 t
        \right]
    - \nonumber \\ & -
        2 \tilde{s}
        \left[
        -   P^2 \left[  69 t^3 + 82 t^2 \tilde{u} + 48 t \tilde{u}^2 + 21 \tilde{u}^3  \right]
        +   24 (p,P)^2 t \left[  2 t + \tilde{u}  \right]
        \right]
    - \nonumber \\ & -
        2 \tilde{s}
        \left[
        -   12 (p,P) (p',P) t \left[  9 t + 8 \tilde{u}  \right]
        +   12 (p',P)^2 t \left[  5 t + 2 \tilde{u}  \right]
        \right]
    + \nonumber \\ & +   
        3 
        \left[ 
            P^2 \left[ t + \tilde{u} \right] 
            \left[  26 t^3 + 20 t^2 \tilde{u} + 3 t \tilde{u}^2 + \tilde{u}^3  \right]
        -   4 (p,P)^2 t \left[  t + \tilde{u}  \right] \left[  9 t + \tilde{u}  \right]
        \right]
    + \nonumber \\ & \left. +   
        3 
        \left[ 
        +   8 (p,P) (p',P) t \left[ t + \tilde{u} \right] \left[  4 t + 5 \tilde{u}  \right]
        -   4 (p',P)^2 t \left[  9 t^2 + 8 t \tilde{u} + \tilde{u}^2  \right]
        \right]
    \right\},
\end{align}
\begin{align} %
    R^{(0)}_{2}
& =
    4 \tilde{s} \tilde{u}
    \left\{
    -   3 \tilde{s}^4 P^2 \tilde{u}
    -   3 \tilde{s}^3
        \left[
            P^2 \left[  t^2 + 3 t \tilde{u} + \tilde{u}^2  \right]
        +   2 (p,P) (p',P) t
        \right]
    + \right. \nonumber \\ & +
        \tilde{s}^2
        \left[
        -   P^2 \left[  3 t + \tilde{u}  \right]
            \left[  4 t^2 + 3 t \tilde{u} + 3 \tilde{u}^2  \right]
        +   6 (p,P)^2 t \left[  t + \tilde{u}  \right]
        \right]
    + \nonumber \\ & +
        \tilde{s}^2
        \left[
        -   6 (p,P) (p',P) t \left[  2 t + \tilde{u}  \right]
        +   6 (p',P)^2 t \left[  2 t + \tilde{u}  \right]
        \right]
    + \nonumber \\ & +
        \tilde{s}
        \left[
            3 t^3 \left[  -5 P^2 t + 6 (p,P)^2 - 2 (p,P) (p',P) + 8 (p',P)^2  \right]
        \right]
    + \nonumber \\ & +
        \tilde{s}
        \left[
        -   4 t^2 \tilde{u} \left[ 5 P^2 t - 3 (p,P)^2 + (p,P) (p',P) - 3 (p',P)^2  \right]
        \right]
    + \nonumber \\ & +
        \tilde{s}
        \left[
            t \tilde{u}^2
            \left[
                6 \left[  (p,P)^2 - (p,P) (p',P) + (p',P)^2  \right]
            -   13 P^2 t
            \right]
        -   9  P^2 t \tilde{u}^3
        -   3 P^2 \tilde{u}^4
        \right]
    - \nonumber \\ & -
        3 t \left[ t + \tilde{u} \right]
        \left[
            t \left[  2 t + \tilde{u}  \right]
            \left[  P^2 \left[  t + \tilde{u}  \right] - 2 (p',P)^2  \right]
        \right]
    - \nonumber \\ & \left. -
        3 t \left[ t + \tilde{u} \right]
        \left[
        -   4 (p,P)^2 t \left[  t + \tilde{u}  \right]
        +   2 (p,P) (p',P) \tilde{u} \left[  t + \tilde{u}  \right]
        \right]
   \right\},
\end{align}
\begin{align} %
    R^{(0)}_{1}
& =
    2 \tilde{s}^2 t^2 \tilde{u}^2
    \left\{
    -   7 \tilde{s}^2 P^2
    +   4 \tilde{s}
        \left[  -5 P^2 t + 6 (p,P)^2 - 6 (p,P) (p',P) + 8 (p',P)^2  \right]
    - \nonumber \right. \\ & -
        14 \tilde{s} P^2 \tilde{u}
    +   4 \tilde{u} \left[  - 5 P^2 t + 8 (p,P)^2 - 6 (p,P) (p',P) + 6(p',P)^2  \right]
    + \nonumber \\ & \left. +
        4 t \left[  - 3 P^2 t + 7 (p,P)^2 - 2(p,P) (p',P) + 7 (p',P)^2  \right]
   -    7 P^2 \tilde{u}^2
   \right\},
\end{align}
\begin{align} %
    R^{(0)}_{0}
& =
    2 \tilde{s}^2 t^2 \tilde{u}^2
    \left\{ \right.
    -   \tilde{s}^3 P^2
    -   \tilde{s}^2
        \left[
            P^2 \left[   t + 3 \tilde{u}  \right]
        +   4 (p',P) \left[  (p',P) - 2(p,P)  \right]
        \right]
   - \nonumber \\ & -
        2 \tilde{s} \tilde{u}
        \left[  P^2 t + 4 \left[  (p,P)^2 - 3 (p,P)(p',P) + (p',P)^2  \right]  \right]
   - \nonumber \\ & -    
        3 \tilde{s} P^2 \tilde{u}^2
   -    4 \tilde{s} t \left[  (p,P) - (p',P)  \right]^2
   - \nonumber \\ & -
        \tilde{u}^2
        \left[
            P^2 \left[  t + \tilde{u}  \right]
            + 4 (p,P) \left[  (p,P) - 2(p',P)  \right]
        \right]
   -
        4 t \tilde{u} \left[  (p,P) - (p',P)  \right]^2
   \left.\right\}.
\end{align}
\begin{align} 
    R^{(2)}_{5}
& =
    4 
    \left\{
        \tilde{s}^2 \left[ 4 (p',P)^2 - P^2 t \right]
    +   2 \tilde{s} \tilde{u} \left[  5 P^2 t + 4 (p,P)(p',P)  \right]
   + \tilde{u}^2 \left[  4 (p,P)^2-P^2t  \right]
   \right\}
\end{align}
\begin{align} 
    R^{(2)}_{4}
& =
    \tilde{s}^3 \left\{  8 P^2 t - 34 P^2 \tilde{u} - 32 (p',P)^2 \right\}
    - \nonumber \\ & -   
        4 \tilde{s}^2 
        \left\{
            P^2 \left[  -2 t^2 + 42 t \tilde{u} - 7 \tilde{u}^2  \right]
        +   16 (p,P) (p',P) \tilde{u}
        +   8 (p',P)^2 \left[  t + \tilde{u}  \right]
        \right\}
    + \nonumber \\ & +    
        2 \tilde{s} \tilde{u} 
        \left\{
            4 t \left[  -22 P^2 t + 17 (p,P)^2 + 6 (p,P)(p',P) + 17 (p',P)^2  \right]
        \right\}
    + \nonumber \\ & +    
        2 \tilde{s} \tilde{u} 
        \left\{
        -   4 \tilde{u} \left[  21 P^2 t + 4 (p,P) \left[  (p,P) + 2 (p',P)  \right]  \right]
        -   17 P^2 \tilde{u}^2
        \right\}
    - \nonumber \\ & -    
        8 \tilde{u}^2 \left\{  t + \tilde{u}  \right\} \left\{  4 (p,P)^2 - P^2 t  \right\}
\end{align}
\begin{align} 
    R^{(2)}_{3}
& =
    -4 
    \left\{ \right.
        \tilde{s}^4 \left[  P^2 \left[  t - 11 \tilde{u}  \right] - 4 (p',P)^2  \right]
    - \nonumber  \\ & -    
        \tilde{s}^3 
        \left[
            P^2 \left[  -2 t^2 + 53 t \tilde{u} + 9 \tilde{u}^2  \right]
        +   8 (p,P) (p',P) \tilde{u}
        +   8 (p',P)^2 \left[ t - 2 \tilde{u} \right]
        \right]
    + \nonumber \\ & +    
        \tilde{s}^2 
        \left[
            P^2 \left[  t + \tilde{u}  \right] \left[  t^2 - 99 t \tilde{u} - 9 \tilde{u}^2  \right]
        +   4 (p,P)^2 \tilde{u} \left[  11 t - \tilde{u}  \right]
        \right]
    + \nonumber \\ & +    
        \tilde{s}^2 
        \left[
            8 (p,P) (p',P) \tilde{u} \left[  7 t+4 \tilde{u}  \right]
        -   4 (p',P)^2 \left[  t^2 - 25 t \tilde{u} + \tilde{u}^2  \right]
        \right]
    + \nonumber \\ & +    
        \tilde{s} \tilde{u}
        \left[
        -   P^2 \left[  58 t^3 + 98 t^2 \tilde{u} + 53 t \tilde{u}^2 + 11 \tilde{u}^3  \right]
        +   4 (p,P)^2 \left[  17 t^2 + 25 t \tilde{u} + 4\tilde{u}^2  \right]
        \right]
    + \nonumber \\ & +    
        \tilde{s} \tilde{u}
        \left[
            8 (p,P) (p',P) \left[  6 t^2 + 7 t \tilde{u} - \tilde{u}^2  \right]
        +   4 (p',P)^2 t \left[  17 t + 11 \tilde{u}  \right]
        \right]
    - \nonumber \\  & -    
        \tilde{u}^2 \left[ t + \tilde{u} \right]^2 \left[  4(p,P)^2 - P^2 t  \right]
    \left. \right\}
\end{align}
\begin{align} 
    R^{(2)}_{2}
& =
    2 \tilde{s} \tilde{u} 
    \left\{ 
        -5 \tilde{s}^4 P^2 + 4 \tilde{s}^3 \left[ -7 P^2 t - 2 P^2 \tilde{u} + 12(p',P)^2  \right]
    + \nonumber \right. \\ & +    
        \tilde{s}^2 
        \left[
            t \left[  -85 P^2 t + 20 (p,P)^2 + 72 (p,P) (p',P) + 196 (p',P)^2  \right]
        -   84 P^2 t \tilde{u}
        \right]
    + \nonumber \\ & +    
        \tilde{s}^2 
        \left[
        -   6 P^2 \tilde{u}^2 
        +   48 (p',P) \tilde{u} \left[  2 (p,P) +(p',P)  \right]
        \right]
    + \nonumber \\ & +    
        2 \tilde{s} 
        \left[
            8 t^2 \left[  -7 P^2 t + 6 (p,P)^2 + 8 (p,P) (p',P) + 14 (p',P)^2  \right]
        \right]
    + \nonumber \\ & +    
        2 \tilde{s} 
        \left[
            t \tilde{u} \left[ -85 P^2 t + 52 (p,P)^2 + 184 (p,P) (p',P) + 52 (p',P)^2  \right]
        \right]
    + \nonumber \\ & +    
        2 \tilde{s} 
        \left[
            6 \tilde{u}^2 \left[  4 (p,P) \left[  (p,P) + 2 (p',P)  \right] - 7 P^2t  \right]
        -   4 P^2 \tilde{u}^3
        \right]
    + \nonumber \\ & +    
        4 t^3 \left[  -12 P^2 t + 17 (p,P)^2 + 14 (p,P) (p',P) + 17 (p',P)^2  \right]
    + \nonumber \\ & +   
        16 t^2 \tilde{u} \left[  -7 P^2 t + 14 (p,P)^2 + 8 (p,P) (p',P) + 6 (p',P)^2  \right]
    + \nonumber \\ & +    
        t \tilde{u}^2 \left[  -85 P^2 t + 196 (p,P)^2 + 72 (p,P) (p',P) + 20 (p',P)^2  \right]
    + \nonumber \\  & \left. +   
        4 \tilde{u}^3 \left[  12 (p,P)^2 - 7 P^2 t  \right]
    -   5 P^2 \tilde{u}^4
    \right\}
\end{align}
\begin{align} 
    R^{(2)}_{1}
& =
    4 \tilde{s} t \tilde{u} 
    \left\{ 
    -   \tilde{s}^4 P^2
    +   \tilde{s}^3 \left[  -3 P^2 t - 4 P^2 \tilde{u} + 8 (p,P) (p',P)  \right]
    + \nonumber \right. \\ & +   
        \tilde{s}^2 
        \left[
            \tilde{u} \left[8 \left[  (p,P)^2 + (p,P) (p',P) + (p',P)^2  \right] - 9 P^2 t  \right]
        \right]
        + \nonumber \\ & +   
        \tilde{s}^2 
        \left[
        -   2 t \left[  P^2 t + 2 \left[ (p,P)^2 - 4 (p,P) (p',P) + (p',P)^2  \right]  \right]
        -   6 P^2 \tilde{u}^2
        \right]
    + \nonumber \\ & +    
        \tilde{s} 
        \left[
            \tilde{u}^2 \left[  8 \left[  (p,P)^2 + (p,P) (p',P) + (p',P)^2  \right] - 9 P^2 t  \right]
        \right]
    + \nonumber \\ & +    
        \tilde{s} 
        \left[
        -   4 t \tilde{u} \left[  P^2 t + 6 (p,P)^2 - 16 (p,P) (p',P) + 6 (p',P)^2  \right]
        \right]
    + \nonumber \\ & +    
        \tilde{s} 
        \left[
        -   4 P^2 \tilde{u}^3
        -   4 t^2 \left[  (p,P) - (p',P)  \right]^2
        \right]
    - \nonumber \\  & \left. -   
        \tilde{u} \left[ t + \tilde{u} \right] 
        \left[  
            2 P^2 t \tilde{u} 
        +   P^2 \tilde{u}^2 
        +   4 t \left[  (p,P) - (p',P)  \right]^2 
        -   8 (p,P) (p',P) \tilde{u}  
        \right]
    \right\}
\end{align}
\begin{align} 
    R^{(4)}_{4}
& =
    -32 
    \left\{
        \tilde{s}^2 \left[ P^2 t - 4 (p',P)^2  \right]
    +   2 \tilde{s} \tilde{u} \left[ P^2 t - 4 (p,P) (p',P) \right]
    +   \tilde{u}^2 \left[  P^2 t - 4(p,P)^2  \right]
   \right\}
\end{align}
\begin{align} 
    R^{(4)}_{3}
& =
    64 \left[  \tilde{s} + t + \tilde{u}  \right] 
    \left\{
        \tilde{s}^2 \left[  P^2 t - 4 (p',P)^2 \right]
   + \nonumber \right. \\ & \left. +    
        2 \tilde{s} \tilde{u} \left[  P^2 t -  4 (p,P) (p',P)  \right]
   +    \tilde{u}^2 \left[  P^2 t - 4 (p,P)^2  \right]
   \right\}
\end{align}
\begin{align} 
    R^{(4)}_{2}
& =
    -32 \left[ \tilde{s} + t + \tilde{u} \right]^2 
    \left\{
        \tilde{s}^2 \left[  P^2 t - 4 (p',P)^2  \right]
    + \nonumber \right. \\ & \left. +   
        2 \tilde{s} \tilde{u} \left[  P^2 t - 4 (p,P) (p',P)  \right]
    +   \tilde{u}^2 \left[  P^2 t - 4(p,P)^2  \right]
    \right\}
\end{align}

\end{widetext}

\bibliography{bibl}

\end{document}